\documentstyle[11pt]{article}
\begin{document}


\centerline{\large\bf Final State Interactions in 
        $D^0 \to K^0 \bar{K^0}$
	\footnote{Supported in part by National Natural Science 
	Foundation of China } } \vspace{1cm}

\centerline{  You-Shan Dai$ \rm ^{a,d}$,
             Dong-Sheng Du$\rm ^{a,b}$,
             Xue-Qian Li$ \rm ^{b,c}$,
             Zheng-Tao Wei$ \rm ^{a}$ and
             Bing-Song Zou $\rm ^{a}$}

\vspace{1cm}

$\rm ^a$ Institute of High Energy Physics, P.O.Box $918(4)$, Beijing
$100039$, China

$\rm ^b$ CCAST(World Laboratory), ~~P.O.Box $8730$,
Beijing $100080$, China

$\rm ^c$ Department of Physics, Nankai University,
Tianjin $300071$, China

$\rm ^d$ Department of Physics, Hangzhou University,
Zhejiang, $310028 $ China

\vspace*{0.3cm}

\begin{center} \begin{minipage}{12cm}

\noindent{\bf Abstract}

It is believed that the production rate of $D^0\rightarrow K^0\bar K^0$ 
is almost solely determined by final state interactions (FSI) and hence
provides an ideal place to test FSI models. Here we examine model
calculations of the contributions from s-channel resonance
$f_J(1710)$ and t-channel exchange to the FSI effects in 
$D^0\to K^0\bar K^0$. The contribution from s-channel $f_0(1710)$ is small. 
For the t-channel FSI evaluation,
we employ the one-particle-exchange (OPE) model and Regge model respectively.
The results from two methods are roughly consistent with each other
and can reproduce the large rate of $D^0\to K^0\bar K^0$ reasonably well.

\end{minipage} 
\end{center}

\vspace{1.5cm}

\newpage

\baselineskip 24pt

\section*{I.  Introduction}

The importance of the final state interactions (FSI) in hadronic 
processes has been recognized for a long time. Recently its applications 
in D and B decays have attracted extensive interests and attentions of 
theorists.  In many decay modes the FSI may play a crucial role.

Here the FSI refers to the soft rescattering processes at hadronic level
\cite{Donoghue}.  Since all FSI processes are concerning 
non-perturbative QCD and cannot be reliably evaluated with any well-established 
theoretical framework, we have to
rely on phenomenological models to analyze  the FSI effects in certain
reactions. The chiral lagrangian approach is proved to be reliable for 
evaluating hadronic processes, but there are too many free parameters 
which are determined by fitting data, so that its applications are much
constrained. Therefore, we have tried
to look for some simplified models which  can give rise to
reasonable estimation of FSI.  

The decay $D^0\rightarrow K^0\bar K^0$ is a very interesting mode. Pham
\cite{Pham} and Lipkin\cite{Lipkin} noticed the important role of FSI 
to this production long time ago.  A
direct $D^0\rightarrow K^0\bar K^0$ can only occur via a W-boson exchange 
based on the quark-diagram analysis\cite{Chau}, and moreover,
since the CKM entries for $c\bar u\rightarrow d\bar d\;{\rm and}\; 
s\bar s$ have opposite signs, the reaction must be proportional to an 
SU(3) breaking, therefore according to the common knowledge obtained by 
studying B and D decays, direct $D^0\rightarrow K^0\bar K^0$ is much 
suppressed than $D^0\rightarrow K^+K^-$. However, the data show that
$B(D^0\rightarrow K^0\bar K^0)\sim (6.5\pm 1.8)\times 10^{-4}$ 
and $B(D^0\rightarrow K^+K^-)\sim (4.27\pm 0.16)\times 10^{-3}$
\cite{DATA}.  Obviously, the $D^0\rightarrow K^0\bar K^0$ is
realized through inelastic final state interactions.  Namely, 
$K^0\bar{K^0}$ is not a direct product of $D^0$ decay, but is secondary 
one from other hadrons (mesons, mainly) which have larger direct 
production rate in $D^0$ decays, via hadronic rescattering. Hence the 
decay $D^0 \to K^0 \bar{K^0}$ provides an ideal place to test model 
calculations of FSI. 

Around $m_D=1.86GeV$ energy region, there is an abundant spectrum of 
resonances. The s-channel resonance contribution can be very important. 
However only those with approprate quantum numbers ($J^{PC}=0^{++}$) 
can contribute to the FSI for $D^0 \to K^0 \bar{K^0}$ mode. According 
to PDG Tables \cite{DATA}, there is only one $0^{++}$ resonance with 
mass sufficiently close to $m_D$, i.e., the $0^{++}$ component in 
$f_J(1710)$.  In this work, we evaluate s-channel 
contribution by only accounting $f_J(1710)$ and the rest is attributed to
the t-channel exchange. For the t-channel exchange, we consider two 
approaches. One is the One-Particle-Exchange (OPE) model, concretely here 
it is the single-meson-exchange, while another is the Regge pole model. 
In fact, the Regge trajectories contain all non-perturbative QCD effects, 
but from another angle, its leading term is exactly the exchange of a meson  
with appropriate quantum numbers. The calculation with the 
single-meson-exchange scenario is obviously much simpler and straightforward. 
Moreover, some
theoretical uncertainties are included in an off-shell form factor which
modifies the effective vertices, therefore can compensate the residue
effects which exist in a precise Regge pole model. This compensation can 
at least be of the same accuracy as the Regge pole model with several free 
parameters. One can trust that the results obtained in the two approaches 
should be qualitatively consistent, even not exactly equal. Our later 
numerical results confirm this allegation.

In Sec.II, we give the  formulations for s-
and t-channel FSI effects. For t-channel case, both the
One-Particle-Exchange 
model and the Regge pole model are used.    
The numerical results and discussion are given in Sec.III. 

\section*{II.  The formulations}

  The direct decay amplitudes of $D^0\rightarrow VV'$ and $D^0\rightarrow
PP'$ where V(V') and P(P') denote vector and pseudoscalar mesons, are
given in many literatures and we will follow the conventions
of \cite{BSW}.

(1) The s-channel resonance contribution.

Even though the spectrum is abundant at $m_D$ region, only the $0^{++}$ 
component of $f_J(1710)$ can make substantial contributions to the s-channel
FSI. However the $0^{++}$ component of $f_J(1710)$ is still not well
determined \cite{DATA}. we use 
the data of $f_0(1710)$ for our later calculations. It is expected that the
brought errors are within  the error tolerance  region of the present data.

To lowest order, the effective coupling of $f_0$ to $VV'$ and $PP'$
(V,P are vector and pseudoscalar) which are concerned here, can be of forms
\begin{eqnarray}
L_I &=& g\phi^+\phi f\hspace{5cm} {\rm PP'f} \\
L_I &=& g'A_{\mu}A^{\mu}f \hspace{4.7cm} {\rm VV'f}.
\end{eqnarray}
With these lagrangians, 
the effective coupling constants $g$ and $g'$ are obtained by fitting 
the branching ratios of $f_0$ to $VV'$ or $PP'$.

The effective weak decay Hamiltonion for our process is:  
\begin{eqnarray}
H_{eff} &=& \frac{G_F}{\sqrt{2}} \{ V_{us}V_{cs}^*
[c_1(\bar{s}c)_{V-A}(\bar{u}s)_{V-A}+ c_2(\bar{s}s)_{V-A}(\bar{u}c)_{V-A}]+
\nonumber\\
& & V_{ud}V_{cd}^* ~~ [c_1(\bar{d}c)_{V-A}(\bar{u}d)_{V-A}+
c_2(\bar{d}d)_{V-A}(\bar{u}c)_{V-A}] \},
\end{eqnarray}
where $V_{us}$,
$V_{cs}$, $V_{ud}$, $V_{cd}$ are the CKM matrix entries, $V-A$ represents
$\gamma_{\mu}(1-\gamma_5)$.  

The amplitude of the decay $D^0\to K^+ K^-$ is:  
\begin{eqnarray} A(D^0\to K^+ K^-) &=& \frac{G_F}
{\sqrt{2}}V_{us}V_{cs}^*a_1 <K^+ K^-|(\bar{s}c)_{V-A}
(\bar{u}s)_{V-A}|D^0> \nonumber\\ 
&=& \frac{G_F}{\sqrt{2}}V_{us}V_{cs}^*a_1 
(-f_K F_0^{DK}(m_K^2)(m_D^2-m_K^2)),
\end{eqnarray} 
where $a_1=c_1+\frac{1}{N_c}c_2$.  The
non-factorization effects are neglected here.

In terms of these effective couplings, the amplitude of $D^0\rightarrow K^0\bar
K^{\; 0}$ with the s-channel resonance $f_0(1710)$ contribution
can be written as
\begin{eqnarray}
\label{s}
A^{FSI} &=& \sum_{all\; MM'}\frac{1}{2} \int \frac{d^3 p_1}
{(2\pi)^3 2E_1} \frac{d^3 p_2}{(2\pi)^3 2E_2}(2\pi)^4
\delta^4(p_1+p_2-p_B) A(D^0\to M+M') \nonumber \\
&& \cdot g_{MM'}\cdot g_{K K}
\frac{i}{s^2-m_{f_0}^2+im_{f_0} \Gamma_{tot}}\times \kappa
\end{eqnarray} 
where $\frac{i}{s^2-m_{f_0}^2+im_{f_0} \Gamma_{tot}}$ is the 
relativistic Breit-Wigner resonance propagator for $f_0(1710)$ and $s$
is the total c.m. energy square and $\kappa$ is an isospin factor.
The sum over the
weak amplitudes $A(D^0\rightarrow MM')$ includes all possible states.
The physical picture is shown in Fig.1. For other intermediate mesons
other than $f_0(1710)$, their propagators may provide a
suppression factor $1/(m_D^2-m^2)$ which would wash out their
contributions.

Thus it is easy to derive
\begin{eqnarray}
&& A(D^0\rightarrow K^+ K^-\rightarrow f_0(1710)\rightarrow K^0\bar
K^{\; 0}) \nonumber \\
&& ={G_F\over\sqrt 2}{g_{KKf}^2\over 32\pi}V_{us}V_{cs}^* a_1 f_K
F_0^{DK}(m_D^2-m_K^2)(1-{4m_K^2\over m_D^2})^{1/2}\times \nonumber\\
&& \frac{(m_D^2-m_f^2)-i\Gamma_fm_f}{(m_D^2-m_f^2)^2+\Gamma_f^2m_f^2}.
\end{eqnarray}
For the vector meson case, we have
\begin{eqnarray}
&& A(D^0\rightarrow \rho^+\rho^-\rightarrow f_0(1710)\rightarrow K^0\bar
K^{\; 0}) \nonumber \\
&& ={G_F\over\sqrt 2}{g_{KKf}\cdot g_{\rho\rho f}\over 16\pi}V_{ud}V_{cd}^*f_K
\sqrt{1\over 3}a_1 (1-{4m_{\rho}^2\over m_D^2})^{1/2} \nonumber\\
&& [m_{\rho}(m_D+m_{\rho})f_{\rho}A_1(2+{(m_D^2-2m_{\rho}^2)^2\over 4m_{\rho}^4})
- {2m_{\rho}\over m_D+m_{\rho}}f_{\rho}A_2(m_D^2-{m_D^4\over 2m_{\rho}^2}
\nonumber \\
&& + {m_D^4(m_D^2-2m_{\rho}^2)\over 8m_{\rho}^4})]\times
\frac{(m_D^2-m_f^2)-i\Gamma_fm_f}{(m_D^2-m_f^2)^2+\Gamma_f^2m_f^2},
\end{eqnarray}
where all $F_0^{DK},a_1,A_1,A_2$ etc. are defined according to the conventions
in \cite{BSW}.

For $A(D^0\rightarrow \pi^+\pi^-\rightarrow f_0(1710)\rightarrow K^0\bar
K^{ 0})$, we need to replace $g^2_{KKf}$, $V_{us}V^*_{cs}$, 
$m_K^2$ in the expression
by $g_{KKf}g_{\pi\pi f}$, $V_{ud}V^*_{cd}$, $m_{\pi}^2$.
For other intermediate states such as $K^{*+}K^{*-}$ and
$\rho^+\rho^-$, we have no data about their branching ratios of
$f_0(1710)$, so we do not consider their s-channel contribution at
present.

(2) t-channel contribution: The OPE model.

In the OPE model, a single t-channel (the same as u-channel) virtual 
particle is exchanged, (see Fig.2) and it is natural to assume that the 
lightest particle with proper quantum number dominates.

The exchange scenario has been studied in $D\rightarrow VP $ and 
$B\to \pi K$ cases \cite{Zou,Wei}.  The effective vertices of strong 
interaction for the rescattering process, such as $g_{KK^*\pi}, 
g_{\rho KK}$ etc.  are gained from data provided the flavor SU(3) 
symmetry holds.  However, since the t-channel exchanged particles P 
and V are off their mass shell, so a phenomenological form factor 
$F(\Lambda)={\Lambda^2-m^2\over \Lambda^2-t}$ is introduced to 
compensate the off-shell effect at the vertices \cite{Locher}.  
It is noted that $\Lambda$ is a parameter which takes value between 
1.2$\sim$2.0 GeV in normal sense.  As pointed out in last section, 
the parameter $\Lambda-$value would smear the errors caused by assuming 
the dominates of one-particle exchange. 

Obviously, for $D^0\rightarrow K^0\bar K^0$ final state, PV intermediate
state is forbidden, meanwhile we also ignore contributions from the
intermediate states with more than two mesons or baryons, which are
definitely much smaller.

There are two key aspects to make the concerned processes substantial.  
First the direct production amplitude of D$\rightarrow$ PP or VV must 
be large enough, and the second, the scattering amplitude of PP (or VV)
$\rightarrow K^0\bar K^0$ is not small. It depends on the effective
couplings and how far the propagating meson deviates from its mass shell.  
Since the scattering PP (or VV)$\rightarrow K^0\bar K^0$ are, in general, 
inelastic processes, the absolute values of the amplitudes are smaller 
than unity.

First, let us study which channels of D$\rightarrow$ PP or VV are
substantially large.  Here let us
just make some order estimations of the amplitudes before doing concrete
calculations.

Based on the quark diagrams, definitely $D^0\rightarrow K^+K^-,\; K^{*+}
K^{*-},\;\pi^+\pi^-,\;\rho^+\rho^-$ have larger amplitudes because they 
are realized via the so-called external W-emission \cite{Chau} which are 
much larger than other mechanisms.  

For $D^0\rightarrow\pi^0\pi^0$ or $\rho^0\rho^0$, even though they can 
happen via the internal W-emission, the amplitudes are about 3 times 
smaller than the external W-emission as
$$\sqrt{{\Gamma(D^0\rightarrow\pi^0\pi^0)\over\Gamma(D^0
\rightarrow\pi^+\pi^-)}}=({a_2\over \sqrt 2 a_1})\sim
0.3\sim\sqrt{{\Gamma(D^0\rightarrow\rho^0\rho^0)
\over\Gamma(D^0\rightarrow\rho^+\rho^-)}}.$$ Therefore, in our later
calculations, we neglect contributions from such intermediate states.

(i) The $D^0\to PP\to K^0\bar{K^0}$ case

Here we present the formulae for $D^0\to K^+ K^- \to K^0\bar{K^0}$ as an
example and a similar expression can be written down for $D^0\to \pi^+ \pi^-
\to K^0\bar{K^0}$.  In this case the exchanged meson is $\rho^{\pm}$.  

It is believed \cite{Zou,Wei} that a single particle exchange in the
t-channel would make dominant contributions to the FSI.  For $D^0\to K^+ K^-
\to K^0 \bar{K^0}$ process shown in Fig.2, $K^+$, $K^-$, and the t-channel
exchanged $\rho$ form a triangle diagram. As a matter of fact,
for a pure FSI process, we only need to evaluate the absorptive part of the
triangle. Definitely,
the dispersive part of this loop
can be calculated in terms of the dispersion relation \cite{Locher},
and generally it is expected to be of the same order as the absorptive part of
the loop.

According to the Cutkosky rule, we make cuts to let $K^+$, $K^-$ be on shell,
leaving $\rho^{\pm}$ to be off shell.  At the $K^+ \rho^{\pm} K^0$ vertex
the effective Hamiltonian is $g_{\rho KK}\epsilon_\mu(p_{K^+}+p_{K^0})^\mu $.

The amplitude of $D^0\to K^+ K^- \to K^0 \bar{K^0}$ is:  
\begin{eqnarray}
A^{FSI}& =& \frac{1}{2} \int \frac{d^3 p_1}{(2\pi)^3 2E_1} 
\frac{d^3 p_2}{(2\pi)^3 2E_2}(2\pi)^4 \delta^4(p_1+p_2-p_B) 
A(D^0\to K^+ K^-) \nonumber\\ 
&& \cdot~ g_{\rho KK}^2(p_1+p_3)^{\mu}(p_2+p_4)^{\nu}
(-g_{\mu\nu}+\frac{k_{\mu}k_{\nu}}{m_{\rho}^2})
\frac{i}{k^2-m_{\rho}^2}F(k^2) \nonumber\\ 
& = &\int_{-1}^{1}d(cos\theta)
\frac{|\stackrel{\rightarrow}{p_1}|}{16\pi m_D} g_{\rho KK}^2
\frac{iF(k^2)}{k^2-m_{\rho}^2}H \cdot A(D^0\to K^+K^-), 
\end{eqnarray} where
$H=-(p_1\cdot p_2+ p_1\cdot p_4+ p_2\cdot p_3+ p_3\cdot p_4)$ and
$A(D^0\rightarrow K^+K^-)$ is the direct weak production amplitude.  
We have set $p_1=p_{_{K^+}}$, $p_2=p_{_{K^-}}$, $p_3=p_{_{K^0}}$,
$p_4=p_{_{\bar{K^0}}}$,  $\theta$ is the angle between
$\stackrel{\rightarrow}{p_1}$ and $\stackrel{\rightarrow}{p_3}$.  ~$k$ is the
4-momentum of exchanged particle $\rho$, and
$k^2=(p_1-p_3)^2=m_1^2+m_3^2-2E_1 E_3+ 2|\stackrel{\rightarrow}{p_1}|
|\stackrel{\rightarrow}{p_3}|cos\theta$.

The form factor $F(k^2)$ in Eq.(14) is an off-shell form factor for the vertices
$\rho KK$.  Because the effective coupling constant $g_{\rho KK}$ is obtained
from the data where the three particles are all on-shell, while in our case the
exchanged $\rho-$meson is off-shell, a compensation form factor is needed.
We take $F(p_{\rho}^2)=(\frac{\Lambda^2-m_{\rho}^2}
{\Lambda^2-p_{\rho}^2})^2$ as suggested in Ref.\cite{Zou},

(ii) $D^0\to VV\to K^0\bar{K^0}$ case

The case for intermediate states of two vector mesons (VV) has been studied
in $B\to \rho K^* \to \pi K$ processes \cite{Wei}.  It is shown that the VV
intermediate states give a significant contribution to final state
interactions.  Here we take $D^0\to K^{*+}K^{*-}\to K^0\bar{K^0}$ as an
example, while the expression for $D^0\to \rho^+ \rho^-\to K^0\bar{K^0}$ is
in close analog.

The amplitude for $D^0\to K^{*+} K^{*-}$ decay is:  
\begin{equation} 
A(D^0\to K^{*+} K^{*-}) =\frac{G_F}{\sqrt{2}}
V_{us}V_{cs}^*a_1\cdot M^{K^{*+}K^{*-}}, 
\end{equation} where 
\begin{eqnarray} 
M^{K^{*+} K^{*-}} &\equiv &<K^{*+}|(\bar{u}s)_{V-A}|0> 
<K^{*-}|(\bar{s}c)_{V-A}|D^0> \nonumber\\ 
& = &m_{K^*}(m_D+m_{K^*})f_{K^*}A_1^{D K^*}(m_{K^*}^2)
(\epsilon_{K^{*+}}\cdot\epsilon_{K^{*-}}) \nonumber\\ 
&&-\frac{2m_{K^{*}}}{m_{D}+m_{K^*}}f_{K^*}A_2^{D K^*}(m_{K^*}^2)
(\epsilon_{K^{*+}}\cdot p_D) (\epsilon_{K^{*-}}\cdot p_D) \nonumber\\ 
&&-i\frac{2m_{K^{*}}}{m_{D}+m_{K^*}}f_{K^*}V^{D K^*}(m_{K^{*}}^2)
\epsilon_{\mu\nu\rho\sigma}\epsilon_{K^{*+}}^{\mu}
\epsilon_{K^{*-}}^{\nu}p_{K^{*+}}^{\rho}p_{K^{*-}}^{\sigma}.  
\end{eqnarray}
Unlike the $D^0 \to K^+ K^- \to K^0\bar{K^0}$, the t-channel exchanged
particle in $D^0\to K^{*+}K^{*-}\to K^0\bar{K^0}$ is $\pi^{\pm}$.  Since it
is the lightest meson of right quantum number, it should give rise to the
largest contribution.  The amplitude for the final state interaction of the
process $D^0\to K^{*+}K^{*-}\to K^0\bar{K^0}$ is :  
\begin{eqnarray} 
A^{FSI}&= & \frac{1}{2}\int \frac{d^3 p_1}{(2\pi)^3 2E_1}\frac{d^3 p_2}
{(2\pi)^3 2E_2}(2\pi)^4 \delta^4(p_1+p_2-p_B) \nonumber\\
 & & \cdot A(D^0 \to K^{*+}K^{*-})
 <K^{*+}K^{*-}|S|K^0\bar{K^0}>\nonumber \\ 
 & =&\int_{-1}^{1}d(cos\theta) \frac{|\stackrel{\rightarrow}
 {p_1}|}{16\pi m_{D}}\frac{iF(p_{\pi^0}^2)}
 {(p_{\pi^0}^2-m_{\pi^0}^2)}[m_{K^{*}}(m_{D}+m_{K^*})
 f_{K^*}A_1^{D K^*}(m_{K^{*}}^2) \cdot H_1\nonumber \\ 
 && -\frac{2m_{K^{*}}}{m_{D}+m_{K^*}}m_{D}^2 f_{K^*}A_2^{D 
K^*}(m_{K^{*}}^2)\cdot H_2], 
\end{eqnarray} 
where S is the S-matrix of strong
interaction, $\theta$ is the angle between $\stackrel{\rightarrow}{p_1}$ and
$\stackrel{\rightarrow}{p_3}$, and 
\begin{eqnarray} 
H_1 &=&4g_{K^*K\pi}^2[(p_3\cdot p_4) -\frac{(p_2\cdot p_3)
(p_2\cdot p_4)}{m_2^2 }\nonumber\\ 
&& -\frac{(p_1\cdot p_3)(p_1\cdot p_4)}{m_1^2 } +\frac{(p_1\cdot
p_2)(p_2\cdot p_3)(p_2\cdot p_4)}{m_1^2 m_2^2}] \nonumber \\ 
H_2 &= &4g_{K^*K\pi}^2[(p_3^0 m_4^0) -\frac{(p_2^0 p_3^0)
(p_2\cdot p_4)}{m_2^2 }\nonumber\\ 
&& -\frac{(p_1^0 p_4^0)(p_1\cdot p_3)}{m_1^2 } +\frac{(p_1^0
p_2^0)(p_2\cdot p_3)(p_2\cdot p_4)}{m_1^2 m_2^2}].  
\end{eqnarray} As
mentioned above, the expression for $A^{FSI}(D^0\rightarrow\rho^+\rho^-
\rightarrow K^0\bar K^0)$ is similar, the only distinction is that for
$D^0\rightarrow\rho^+\rho^-\rightarrow K^0\bar K^0$, the exchanged 
particle is $K^{\pm}$ instead.

(3) t-channel contribution: The Regge pole model

The principles of Regge theory are \cite{Collins}:  (i) The scattering
amplitudes are analytic functions of the angular momentum $J$; (ii) A
particle of mass $m$ and spin $\sigma$ will be on a Regge trajectory 
$\alpha(t)$ (where $t$ is the Mandelstam invariant parameters) and 
$\sigma=\alpha(m^2)$; (iii) The partial wave amplitude has a pole of 
the form $\frac{1}{J-\alpha(t)}$.  It is suggested that the Regge 
theory provides a very simple and economical description of total cross 
section at high energy region\cite{Landshoff}.

The invariant amplitude for the scattering of particles with helicities
${\lambda_i}$ from the Regge phenomenology is \cite{Collins}:  
\begin{eqnarray} {\cal M}_{i\rightarrow f}^{\lambda_1 \lambda_2; 
\lambda_3 \lambda_4}= -\left(\frac{-t}{s_0} \right)^{\lambda/2}~
\frac{e^{-i \pi \alpha(t)} + {\cal J}} {2 \sin\pi \alpha(t)} 
\gamma_{\lambda_3 \lambda_4}^{\lambda_1 \lambda_2}
\left( {s\over s_0} \right)^{\alpha (t)}\ \  
\end{eqnarray} 
where $s$ and $t$ are Mandelstam invariants, and
$\lambda=|\lambda_3-\lambda_1|+|\lambda_4-\lambda_2|$.
$\cal J$ is the signature for Regge trajectory.  For Pomeron and $\pi$
trajectory, $\cal J$$= +1$; For $\rho$ and $K^*$ trajectory, 
$\cal J$$= -1$.  This expression corresponds to an asymptotic behavior 
when $s\gg s_0$ and $s_0$ is a scale parameter.  In most
literatures, $s_0$ is taken as $1\;{\rm GeV}^2$.  This Regge asymptotic 
behavior works very well in the energy region $\sqrt s \ge 5GeV$.  
We extend the energy region to $\sqrt s=m_{_D}$.
The legitimacy is likely because we have accounted the s-channel resonance
$f_0(1710)$ contribution separately, while contributions from
the rest resonances can be treated as a smooth function of $s$ which is
determined by the crossed t-channel exchange \cite{Gerard} and it is the
fundamental of the Regge pole theory. Thus we can
assume that there
would not be a large deviation from the Regge asymptotic behavior.
$\gamma (t)$ is a residue function. The linear Regge trajectory as 
an approximation is taken for our calculations 
$\alpha (t)=\alpha_0+\alpha't$.  $\alpha'$ is nearly a universal 
parameter for all Regge trajectories (except for Pomeron), 
$\alpha' \approx 0.9$.  $\alpha_0=0.5$ for $\rho$ and $\omega$ 
trajectories;
$\alpha_0=0.3$ for $K^*$ trajectory; 
$\alpha_0=0$ for $\pi$ trajectories;
$\alpha_0=-0.3$ for $K$ trajectories.  
But in our calculation we have adopt approximation $\alpha_0=0.5$ 
for $\rho$ and $K^*$ trajectories; $\alpha_0=0$
for $\pi$ and $K$ trajectories in order to carry out dispersion 
integration analytically.

We take the $D^0\to K^{*+}K^{*-}\to K^0 \bar{K^0}$ as an example and for
the other intermediate states expressions are similar.

First, we rewrite the helicity amplitude of $D^0\to VV$ decay
in a convenient form \cite{Golowich}:
\begin{eqnarray}
A_{\lambda_1\lambda_2} & =&
<V_1(k_1,\lambda_1)V_2(k_2,\lambda_2)| H_w|D^0(p)>\nonumber \\ 
&=&\epsilon_{\mu}^*(k_1,\lambda_1)\epsilon_{\nu}^*(k_2,\lambda_2)
[ag^{\mu\nu}+\frac{b}{m_1m_2} p^{\mu}p^{\nu}+i\frac{c}{m_1m_2} 
\epsilon^{\mu\nu \alpha \beta}k_{1\alpha}p_{\beta} ] 
\end{eqnarray} 
where $\lambda_1$, $\lambda_2$ are the helicity of $V_1$, $V_2$, 
and $\epsilon_{\mu}$, $\epsilon_{\nu}$ are the polarization vector 
of $V_1$, $V_2$.  From Eq.(10), the above factors for decay
$D^0 \to K^{*+} K^{*-}$ are:  
\begin{eqnarray}
a&=&\frac{G_F}{\sqrt{2}}V_{us}^* V_{cs}a_1 (m_D+m_{K^*})m_{K^*}
f_{K^*}A_1^{D K^*}(m_{K^*}^2) \nonumber \\
b&=&-\frac{G_F}{\sqrt{2}}V_{us}^* V_{cs}a_1 \frac{2m_{K^*}^3}
{(m_D+m_{K^*})}f_{K^*}A_2^{D K^*}(m_{K^*}^2) \\ 
c&=&-\frac{G_F}{\sqrt{2}}V_{us}^*V_{cs}a_1 \frac{2m_{K^*}^3}
{(m_D+m_{K^*})} f_{K^*}V^{D K^*}(m_{K^*}^2)\nonumber 
\end{eqnarray}

In the rest frame of the $D^0$, $K^{*+}$ and $K^{*-}$ have the same helicity. 
According to \cite{Golowich}, there are three independent helicity amplitudes:
\begin{eqnarray} 
A_{++}&=&-a+\sqrt{x^2-1}c \nonumber \\
A_{--}&=&-a-\sqrt{x^2-1}c\\ 
A_{00}&=&-xa-(x^2-1)b \nonumber 
\end{eqnarray}
where $x\equiv\frac{k_1k_2}{m_{K^*}^2}=\frac{m_D^2-2m_{K^*}^2}{2m_{K^*}^2}$.

The discontinuity of amplitude for the final state interaction of 
$D^0\to K^{*+}K^{*-} \to K^0 \bar{K^0}$ is:  
\begin{eqnarray} 
{\rm Disc}A^{FSI}& =&\frac{1}{2} \int \frac{d^3 p_1}
{(2\pi)^3 2E_1} \frac{d^3 p_2}{(2\pi)^3
2E_2}(2\pi)^4 \delta^4(p_1+p_2-p_B) \nonumber\\ 
&& \cdot~A(D^0\to K^{*+} K^{*-})_{\lambda\lambda} 
{\cal M}^{\lambda\lambda;00}(K^{*+}K^{*-}\to K^0 \bar{K^0}).  
\end{eqnarray} 
where $\lambda$ is the helicity of the intermediate state $K^*$. 
The discontinuity of this amplitude precisely corresponds to the absorptive 
part of the hadronic triangle (see Fig.2) for the one-particle-exchange case 
where $K^{*+}, K^{*-}$ are on-shell.  For the rescattering of 
$K^{*+} K^{*-}\to K^0 \bar{K^0}$, the exchange trajectory is $\pi$.  
The helicity amplitude $A_{++}$, $A_{--}$, $A_{00}$ all contribute 
to the same helicity state $\{00\}$ of $K^0 \bar{K^0}$.  
\begin{eqnarray} 
{\rm Disc}A^{FSI}&=&\frac{\sqrt{1-\frac{4m_{K^*}^2}{m_D^2}}}{16\pi s}
\int_{t_{min}}^{t_{max}}dt (A_{++}{\cal M}^{++;00}+A_{--}
{\cal M}^{--;00}+A_{00}{\cal M}^{00;00}) \nonumber\\
&=&\epsilon_{\pi}(\frac{s}{s_0})^{\alpha_0-1} 
\end{eqnarray}
where $\epsilon_{\pi}$ represents the value except the factor 
$(\frac{s}{s_0})^{\alpha_0-1}$ and is calculated numerically. 

The full amplitude of the final state interaction of 
$D^0\to K^{*+}K^{*-} \to K^0 \bar{K^0}$ can be obtained by using the
dispersion relation.
\begin{equation}
A^{FSI}=\frac{\epsilon_{\pi}}
{\pi}\int_{4m_{_{K^*}}^2}^{\infty}{ds}
\frac{(\frac{s}{s_0})^{\alpha_0 -1}}{(s-m_{_{D}}^2)} 
=\frac{\epsilon_{\pi}}{\pi m_{_{D}}^2}ln(1-\frac{m_{_{D}}^2}
{4m_{_{K^*}}^2}) 
\end{equation}

For the process $D^0 \to \rho^+ \rho^- \to K^0 \bar{K^0} $, the leading
trajectory of rescattering is $K$.  For the process 
$D^0 \to K^+ K^- \to K^0 \bar{K^0} $ and 
$D^0 \to \pi^+ \pi^- \to K^0 \bar{K^0}$, the leading trajectory of 
rescattering are $\rho$ and $K^*$ respectively.  

\section*{ III.  Numerical results and discussion}

To reproduce the experimental data 
$B(D^0\to K^0\bar K^0)\sim 6.5\times 10^{-4}$, we need the amplitude to be 
$|A(D^0\to K^0\bar K^0)|\sim 3.35\times 10^{-7}GeV$. Now we examine 
numerical results of various FSI amplitudes.

For the s-channel contribution, we take the experimental data\cite{DATA}
as inputs:
$ m_f=1710 {\rm MeV}$; ~~~ $\Gamma_{tot}=133\pm 14 {\rm MeV}$;
$B(K\bar K)=\Gamma_{K\bar K}/\Gamma_{tot}=0.38$; 
$B(\pi \pi)=\Gamma_{\pi \pi}/\Gamma_{K\bar K}=0.39$.
Since other channels of $f_0(1710)$ decays have not been measured yet, we
do not include their contribution in this numerical estimation.
We expect that they will give similar contribution as $K^+K^-$ and 
$\pi^+\pi^-$ modes.
The numerical results of the s-channel $f_0(1710)$ contributions are tabulated
in Table.1.

\begin{table}[hbt]
\caption{FSI amplitudes from s-channel contribution of $f_0(1710)$.}
\begin{center}
\begin{tabular}{|c|c|} \hline \hline
Decay Mode & $A^{FSI}(\rm GeV) $ \\\hline
$D^0 \to K^+ K^- \to K^0 \bar{K^0}$ &
$(-0.24-i0.53)\times 10^{-7}$ \\\hline
$D^0 \to \pi^+ \pi^- \to K^0 \bar{K^0}$ &
$(0.13+i0.32)\times 10^{-7}$ \\ \hline
total & $(-0.11-i0.21)\times 10^{-7}$ \\ \hline
\end{tabular} 
\end{center}
\end{table}

One can notice that contributions from the $K^+K^-$ and $\pi^+\pi^-$
intermediate states interfere destructively, because $V_{cd}\approx
-V_{us}$. The sum of two contributions is small compared with what
experimental data needs. However if parameters of $f_0(1710)$ 
change\cite{Bugg}, the s-channel contributions may become more
important. For more precise estimation, we shall wait for further
experimental information on $0^{++}$ resonances near $M_D$.

For the OPE model and the Regge pole model, we take
$c_1=1.26$, $c_2=-0.51$\cite{Zou}; 
decay constants \cite{DATA,Guo}:  $ f_{\pi}=0.13GeV,~ f_K=0.16GeV, ~
f_{\rho}=0.221GeV, ~ f_{K^*}=0.221GeV; $
and form factors \cite{BSW,Guo}:  $$ \begin{array}{lll}
F_0^{D\pi}(0)=0.692,& A_1^{D\rho}(0)=0.775,& A_2^{D\rho}(0)=0.923,\\
F_0^{DK}(0)=0.762,& A_1^{D K^*}(0)=0.880,& A_2^{D K^*}(0)=1.147.
\end{array} $$

For the OPE model, the effective strong coupling constants are given in
\cite{Zou}: $g_{K^*K\pi}=5.8$ and $g_{\rho\pi\pi}=\sqrt{2}g_{\rho KK}=6.1$.
The parameter $\Lambda$ in the off-shell form factor $F(k^2)$ varies in 
a range of 1.2 to 2.0 GeV\cite{Locher}. In Table 2, we tabulate the 
results corresponding to three cases: 
$\Lambda=1.2$ GeV, $\Lambda=1.6$ GeV, $\Lambda=2.0$ GeV.

\begin{table}[hbt]
\caption{FSI amplitudes from t-channel contributions in the OPE model. }
\begin{center} 
\begin{tabular}{|c|c|c|c|} \hline \hline
Decay Mode & \multicolumn{3}{|c|} { $A^{FSI}(\rm GeV)$ } \\\cline{2-4} &
$\Lambda=1.2{\rm GeV}$ & $\Lambda=1.6{\rm GeV}$ 
& $\Lambda=2.0{\rm GeV}$ \\ \hline 
$D^0 \to K^+ K^- \to K^0 \bar{K^0}$ & $-i1.52\times 10^{-7} $ &
$-i3.23\times 10^{-7} $ & $-i4.52\times 10^{-7} $ \\ \hline 
$D^0 \to \pi^+ \pi^- \to K^0 \bar{K^0}$ & 
$i1.02\times 10^{-7} $ & $i3.11\times 10^{-7} $ &
$i4.89\times 10^{-7} $ \\ \hline 
$D^0 \to K^{*+} K^{*-} \to K^0 \bar{K^0}$ &
$i4.37\times 10^{-7}$ & $i5.89\times 10^{-7} $ & 
$i6.91\times 10^{-7} $ \\\hline 
$D^0 \to \rho^+ \rho^- \to K^0 \bar{K^0}$ & 
$-i1.79\times 10^{-7} $ &
$-i3.02\times 10^{-7} $ & $-i3.98\times 10^{-7} $ \\ \hline 
total & $i2.08\times 10^{-7}$ & $i2.75\times 10^{-7}$ & 
$i3.30\times 10^{-7}$ \\ \hline
\end{tabular}
\end{center}
\end{table}

Here three points are worthy of note. 
(1) The process $D^0 \to K^{*+}K^{*-} \to K^0 \bar{K^0}$ 
has the largest contribution.  The reason is because the 
exchanged particle is the lightest meson, the pion.  This conclusion 
is the same as in \cite{Zou,Wei}.
(2) The predicted amplitude of process 
$D^0 \to K^{*+}K^{*-} \to K^0 \bar{K^0}$ is not very sensitive to the choice 
of the parameter $\Lambda$.  By contrary, for the other three processes, 
the amplitudes are more sensitive to the choice.
As well-known, the OPE model is more applicable
when the virtual exchanged particle is close to its mass shell.  
In fact, the heavier the particle is, or the further it is off its 
mass shell, then the more sensitive the amplitude is to the parameter 
$\Lambda$.
(3) We only calculate the absorptive part which gives imaginary part
of the FSI amplitudes only. It gives the correct
order of magnitude of the FSI effects. The dispersive real part of the
FSI amplitudes can be calculated in terms of the dispersion 
relation\cite{Locher} with additional parameters, and generally 
it is of the same order of magnitude of the absorptive part.
So the OPE model can reproduce the FSI effects rather well.

For the Regge pole model, two different treatment of the residue function
$\gamma(t)$ are assumed. Model I assumes constant 
residue functions $\gamma$\cite{Nir}:
$\gamma_{\pi\pi\rho}^2=\sqrt{2}\gamma_{KK\rho}
=\sqrt{s_0}\cdot \frac{2Y_{\pi p}^2}
{Y_{pp}}\approx 72$;
Model II takes $\gamma(t)$ to make Eq.(13) to be the same as in the OPE
model for t near the mass squared of the leading exchanged particle.
The numerical results are listed in Table 3.

\begin{table}[hbt]
\caption{FSI amplitudes in Regge pole models.}
\begin{center}
\begin{tabular}{|c|c|c|} \hline \hline 
Decay Mode & \multicolumn{2}{|c|} {$A^{FSI}(\rm GeV)$} \\\cline{2-3}
 & Model I & Model II \\ \hline 
$D^0 \to K^+ K^- \to K^0 \bar{K^0}$ & 
$(-0.31-i2.61)\times 10^{-7}$ & $(-1.06-i2.18)\times 10^{-7}$\\\hline 
$D^0 \to \pi^+ \pi^- \to K^0 \bar{K^0}$ & 
$(0.38+i3.17)\times 10^{-7}$ & $(-1.47+i2.38)\times 10^{-7}$ \\ \hline  
$D^0 \to K^{*+} K^{*-} \to K^0 \bar{K^0}$ &
$(-1.13+i0.07)\times 10^{-7}$ & $(-4.08+i2.75)\times 10^{-7}$ \\ \hline  
$D^0 \to \rho^+ \rho^- \to K^0 \bar{K^0}$ & 
$(1.0-i0.67)\times 10^{-7}$ & $(2.80-i1.12)\times 10^{-7}$ \\  \hline
total & $(-0.06-i0.04)\times 10^{-7}$ & $(-3.81+i1.83)\times 10^{-7}$ \\\hline 
\end{tabular} 
\end{center}
\end{table}

Comparing the imaginary part in Regge pole models with the OPE results in
Table 2, the biggist difference is for $D\to VV\to PP$ modes in Model I.
Model I is the conventional approximation of Regge model for high energies.
It is obviously not a good approximation for the $M_D$ energy region.
We found that the main problem is: the high energy approximation
for the t-dependent couplings, $({-t\over s_0})^{\lambda/2}$ in Eq.(13), 
is not good for $VV\to PP$ at the $M_D$ energy. If we replace the
$({-t\over s_0})^{\lambda/2}\gamma(t)$ in Eq.(13) by the corresponding
effective couplings in the OPE model, then we get Model II which gives
results roughly consistent with OPE results.

In summary, for the t-channel FSI contributions to $D^0\to K^0\bar K^0$,
the OPE model and Regge pole model with a proper treatment of $\gamma(t)$
(Model II) are roughly consistent with each other and can reproduce the
experimental data reasonably well. They may be used to estimate t-channel
FSI effects for other D decay channels. The Regge pole model assuming a
constant $\gamma(t)$ (Model I) is not suitable for $M_D$ energy region.
The s-channel FSI contribution from known $0^{++}$ resonances is small. 

The situation of FSI for B meson decays should be different.
There is an s-dependent suppression factor
$(\frac{s}{s_0})^{\alpha(t)}$ in Regge pole model.
The discontinuity of the final state interaction amplitude 
is proportional to $(\frac{s}{s_0})^{\alpha_0-1}$. For inelastic 
rescattering which the exchange trajectory $\alpha_0<1$, the discontinuity 
of the final state interaction amplitude decreases as the energy increases.  
This predicts that the final state interaction will be small in high
energy region. There is no such s-dependent suppression factor in OPE model.
At high energies, the t-channel exchange of heavier particles will become
more important. The s-independent off-shell form factors in OPE model
may be not enough to compensate these effects.
We will continue our study in
$B-$region and the results will be published elsewhere.

\section*{Acknowledgment}

This work is supported in part by National Natural Science Foundation of
China and the Grant of State Commission of Science and Technology of China.

\newpage
\section*{Figure Captions}

Fig. 1. The s-channel resonance FSI contribution. $j$ represents the 
intermediate states.


\begin{thebibliography}{99}

\bibitem{Donoghue} J.F.Donoghue, E.Glowich, A.A.Petrov and J.M.Soares, 
    Phys.Rev.Lett{\bf 77}(1996) 2178.

\bibitem{Pham} X.-Y.Pham, Phys.Lett.  B{\bf 193} (1987), 331.

\bibitem{Lipkin} H.  Lipkin, Phys.Rev.Lett.  {\bf 44} (1980) 710.

\bibitem{Chau} L.-L.  Chau, Phys.Rev.Lett.  {\bf 95} (1983) 1.

\bibitem{DATA} Particle Date Group, Eur.Phys.J.C {\bf 3} (1998) 1.

\bibitem{BSW} M.Wirbel, B.Stech and M.Bauer, Z.Phys.C {\bf 29} (1985)
  637; Z.Phys.C {\bf 34} (1987) 103.

\bibitem{Zou} X.Li and B.Zou, Phys.Lett.  B{\bf 399} (1997), 297.

\bibitem{Wei} D.Du, X.Li, Z.Wei and B.Zou,  Eur.Phys.J. A{\bf 4} (1999) 91.

\bibitem{Locher} O.Gortchakov, M.P.Locher, V.E.  Markuskin and
   S.von Rotz, Z.Phys.A{\bf 353} (1996) 447;\\ A.  Anisovich and E.  Klempt,
   Z.Phys.A{\bf 354} (1996) 197.

\bibitem{Collins} P.D.B.  Collins, Introduction to Regge Theory and High
   Enrgy Physics.  (Cambridge University Press, Cambridge, England, 1977)

\bibitem{Landshoff} A.  Donnachie and P.V.Landshoff, Phys.Lett.  B{\bf 296}
     (1992), 227

\bibitem{Gerard} J.-M.Gerard, J.Pestieau and J.Weyers, 
    Phys.Lett.B {\bf 436} (1998) 363.

\bibitem{Golowich} E.Golowich, S.Pakvasa, Phys.Rev.D {\bf 51} (1995) 1215.

\bibitem{Bugg} D.V.Bugg et al., Phys.Lett.B {\bf 353} (1995) 378;
     BES Collaboration, J.Z.Bai et al., Phys.Rev.Lett. {\bf 77} (1996) 3959.

\bibitem{Guo} D.Du and L.Guo, Z.Phys.C {\bf 75} (1997) 9-15.

\bibitem{Nir} A.  Falk, A.  Kagan, Y.  Nir and A.  Petrov, Phys.Rev.D{\bf 57}
    (1998) 4290-4300,






\end{thebibliography}
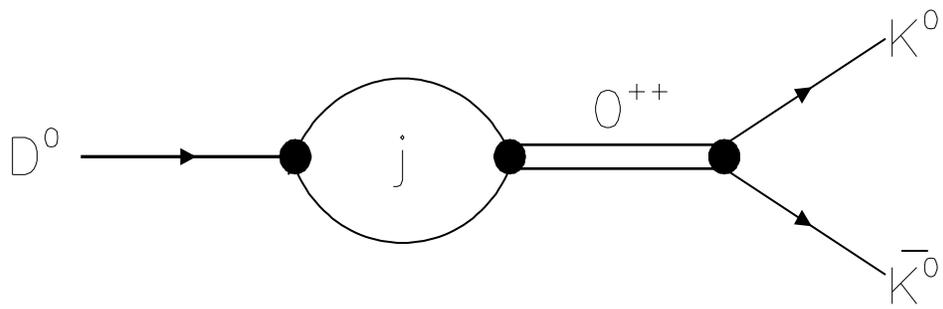
\end{document}